\documentclass[10pt,aps,prl,preprintnumbers,superscriptaddress,showpacs,
twocolumn,floatfix,nofootinbib]{revtex4-1}
\usepackage{graphicx}
\usepackage{amsmath}
\usepackage{slashed}

\newcommand{\gsim}{\;\rlap{\lower 3.5 pt \hbox{$\mathchar \sim$}} \raise 1pt
\hbox {$>$}\;}
\newcommand{\lsim}{\;\rlap{\lower 3.5 pt \hbox{$\mathchar \sim$}} \raise 1pt
\hbox {$<$}\;}

\begin{document}

\title{
\boldmath Regge Limit of Gauge Theory Amplitudes beyond Leading Power Approximation.
\unboldmath}
\author{Alexander A. Penin}
\email[]{penin@phys.ethz.ch}
\affiliation{Department of Physics, University of Alberta, Edmonton, Alberta T6G
2J1, Canada}
\affiliation{Institute for Theoretical Physics, ETH Z\"urich, 8093 Z\"urich,
Switzerland}
\begin{abstract}
We study the high-energy small-angle {\it Regge} limit of the
fermion-antifermion  scattering in  gauge theories and consider the part  of the
amplitude suppressed by a power of the scattering angle.  For abelian gauge
group all-order resummation of the double-logarithmic radiative corrections to
the leading power-suppressed term is performed.  We find that when the logarithm
of the scattering angle is  comparable to the inverse gauge coupling constant
the asymptotic double-logarithmic enhancement overcomes the power suppression, a
formally subleading term becomes dominant, and the small-angle expansion breaks
down. For the nonabelian gauge group we show that in  the color-singlet channel
for sufficiently small scattering angles the power-suppressed contribution
becomes comparable to the one of BFKL pomeron. Possible role of the
subleading-power effects for the solution of the unitarity problem of
perturbative Regge analysis in QED and QCD is discussed. An intriguing  relation
between  the asymptotic behavior of the power-suppressed amplitudes in Regge and
Sudakov  limits is discovered.
\end{abstract}
\preprint{ALBERTA-THY-08-19}

\maketitle
Small-angle or {\em Regge} limit  of high-energy scattering describes a
kinematical configuration with vanishing  ratio of the characteristic
momentum transfer to the total energy of the process.  The asymptotic
behavior   of the gauge theory amplitudes in Regge limit  remains in the
focus of the theoretical studies since the early days of QED and QCD
\cite{Chang:1969by,Cheng:1970jk,Cheng:1970xm,Frolov:1970ij,Lipatov:1976zz,Kuraev:1977fs,Balitsky:1978ic}.
Despite a crucial simplification due to decoupling of the light-cone and
transversal degrees of freedom, the gauge interactions in this limit possess
highly nontrivial dynamics giving a rigorous quantum field theory realization
of Regge concept for high-energy scattering \cite{Collins:2009}. Major
progress has been achieved in the analysis of the leading-power amplitudes
which scale  at small momentum transfer as a ratio of the Mandelstam
variables $s/t\sim 1/\theta^2$, where $\theta$ is the scattering angle. It
culminated in the evaluation  of the next-to-leading QCD corrections to  the
theory of  BFKL pomeron \cite{Fadin:1998py}. At the same time very little is
known about the asymptotic behavior of the  amplitudes suppressed by a small
ratio $\tau=|t/s|$. In contrast to the leading-power amplitudes it is
determined by the double-logarithmic radiative corrections which include the
second power of the large logarithm $\ln\tau$ per each power of the gauge
coupling constant $\alpha$. This type of corrections has been discussed so
far in QED only for the amplitudes   which do not have the leading-power
contribution and   remain finite in the small-angle limit, such as the
electron-to-muon pair forward annihilation \cite{Gorshkov:1966ht}. However
for the general scattering case the logarithmically-enhanced power-suppressed
contributions can also play a crucial role.  Indeed a recent study of the
mass-suppressed amplitudes in the high-energy fixed-angle {\it Sudakov} limit
\cite{Penin:2014msa,Melnikov:2016emg,Penin:2016wiw,Liu:2017vkm,Liu:2018czl}
revealed  that for some processes the double-logarithmic corrections result
in strong enhancement of the power-suppressed terms  which asymptotically
makes them comparable to the leading-power contributions, {\it i.e.} formally
lead to breakdown of the small-mass expansion. If a similar scenario is
realised in the small-angle scattering, it may significantly alter our
understanding of the gauge theory dynamics in Regge limit. Thus a systematic
renormalization group analysis of subleading-power amplitudes is of primary
theoretical interest.

In this Letter we  make the first step toward this goal and discuss the
double-logarithmic behaviour of the leading power-suppressed contribution to
the fermion-antifermion  scattering amplitude. First we consider an abelian
gauge theory and set up an effective theory framework for the analysis of the
Born scattering  in Regge limit. We apply it  to the calculation of the
one-loop power-suppressed double-logarithmic contribution and then perform
the resummation of the double-logarithmic corrections to all orders of
perturbation theory to find the asymptotic behavior of the amplitude. Finally
we discuss the qualitative features  and the impact of the power-suppressed
terms in the  theory of Regge limit in QED and QCD.

We start with the scattering of  a fermion with the initial momentum $p$ and
the final momentum $p'=p+q$ by an external abelian field $A_\mu$ in the limit
$q\to 0$. We choose the reference frame in a such way that the incoming
fermion momentum $p^\mu=(\varepsilon,-\varepsilon,0,0)$ has only one
light-cone component $p^-= p_+=\sqrt{2}\varepsilon$, while $p^+= p_-=0$. In
the high-energy limit one can neglect the fermion mass and the on-shell
conditions $p^2={p'}^2=0$ imply the following scaling of the momentum transfer
components $q_\pm = {\cal O}(\tau)$, $q_\perp = {\cal O}(\tau^{1/2})$, with
$\tau=-q^2/(2\varepsilon)^2$ and $q^2=2q_+q_--q_\perp^2=t$. By expanding a
solution of the free Dirac equation $\psi(p')$ in $q$ we obtain the following
series for the scattering amplitude
\begin{equation}
A^\mu j_\mu= A_- j_++ {\tilde{F}_{+-}\over 4p_+} j^5_+
+{i{F}_{+-}\over 4p_+} j_++{\cal O}(\tau^{3/2})\,,
\label{eq::currentexp}
\end{equation}
where   the currents read
\begin{equation}
\begin{split}
&j_\mu=g\bar\psi(p')\gamma_\mu \psi(p)\,,\\
&j_+=g\bar\psi(p)\gamma_+ \psi(p)\,, \\
&j^5_+=g\bar\psi(p)\gamma_+\gamma_5 \psi(p)\,,
\label{eq::currentdef}
\end{split}
\end{equation}
and $g$ is the fermion charge. In Eq.~(\ref{eq::currentexp}) ${F}_{\mu\nu}$
and $\tilde{F}_{\mu\nu}$ are the gauge field strength tensor and its dual,
respectively. They correspond to the electric and magnetic fields aligned
with the direction of the initial fermion motion. From  the scaling of the
momentum $q$ components  we find $\tilde{F}_{+-}={\cal O}(\tau^{1/2})$ and
${F}_{+-}={\cal O}(\tau)$. Note that the power suppressed terms in
Eq.~(\ref{eq::currentexp}) are quite similar to the ${\cal O}(v)$ Pauli and
${\cal O}(v^2)$ Darwin interaction of a nonrelativistic massive fermion to
the magnetic and electric fields resulting from the expansion of the Dirac
Lagrangian in  small fermion velocity $v$. We can now apply
Eq.~(\ref{eq::currentexp}) for the analysis of the fermion-antifermion
scattering
\begin{equation}
\psi(p_1)+\bar\psi(p_2)\to \psi(p_4)+\bar\psi(p_3)\,
\label{eq::scat}
\end{equation}
with  the initial momenta $p_1^\mu=(\varepsilon,-\varepsilon,0,0)$ and
$p_2^\mu=-(\varepsilon,\varepsilon,0,0)$.
The corresponding Born amplitude  reads
\begin{equation}
{\cal M}^{(0)}={\cal M}^v+{\tau\over 2}\left( {\cal M}^v+{\cal M}^a\right)
+{\cal O}(\tau^2)
\label{eq::Born}
\end{equation}
where the vector and axial component  are defined as follows
\begin{equation}
t{\cal M}^v=j_+(p_1)j_-(p_2),\qquad t{\cal M}^a=j^5_+(p_1)j^5_-(p_2)\,.
\label{eq::ampdef}
\end{equation}
In close analogy with the nonrelativistic expansion, the leading order
amplitude is mediated by a static potential\footnote{In the nonrelativistic
limit  the static Coulomb potential is three-dimensional while  in  Regge
limit it results from the exchange of the {\it Glauber} gluons \cite{Glauber}
propagating in the two-dimensional transversal space.} and  the power
corrections describe a contact fermion-antifermion interaction. Note that if
we perform the expansion in $u$-channel rather than in $s$-channel {\it i.e.}
about the momentum $p_3$ instead of $p_2$, the sign of the axial term
in Eq.~(\ref{eq::Born}) changes.

\begin{figure}[t]
\begin{center}
\begin{tabular}{ccc}
\includegraphics[width=1.8cm]{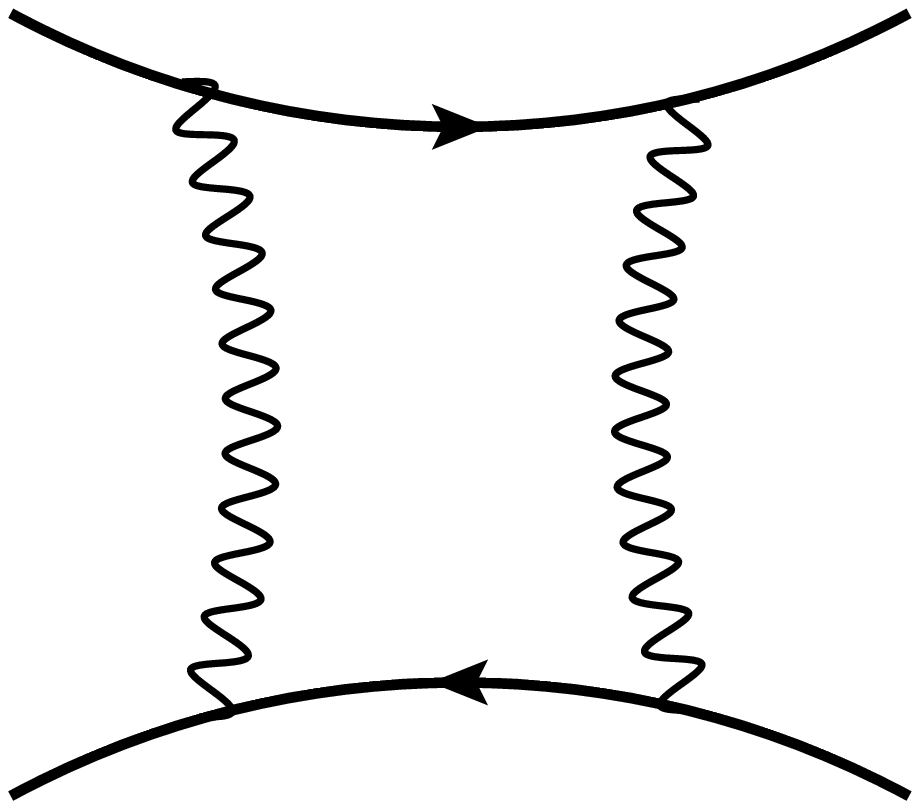}&
\hspace*{05mm}\includegraphics[width=1.8cm]{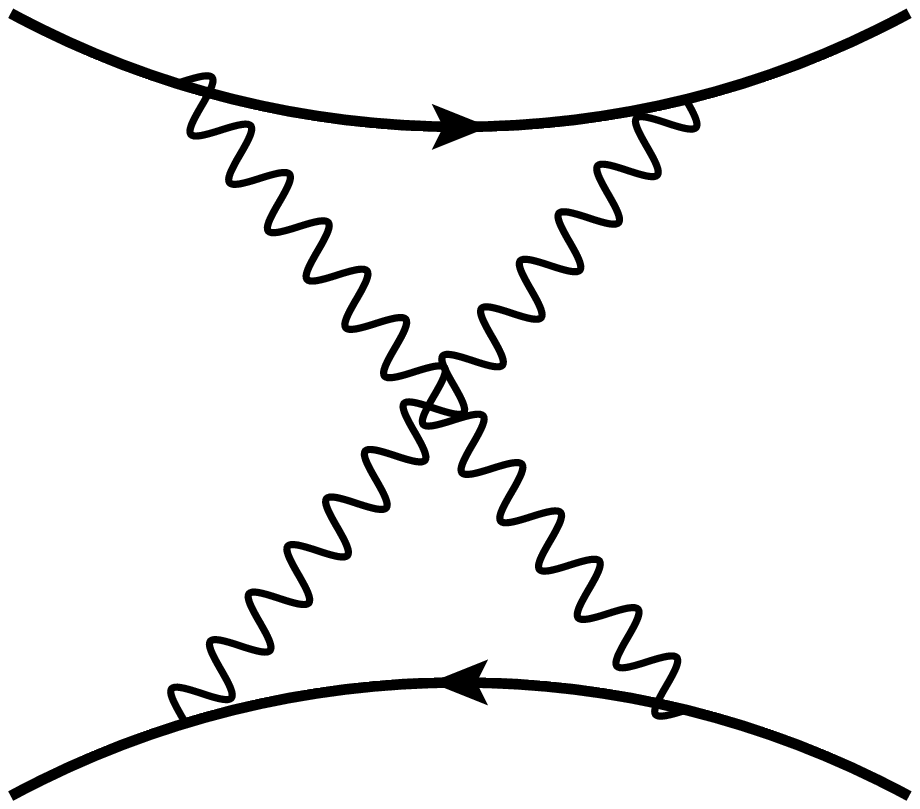}&
\hspace*{05mm}\includegraphics[width=1.8cm]{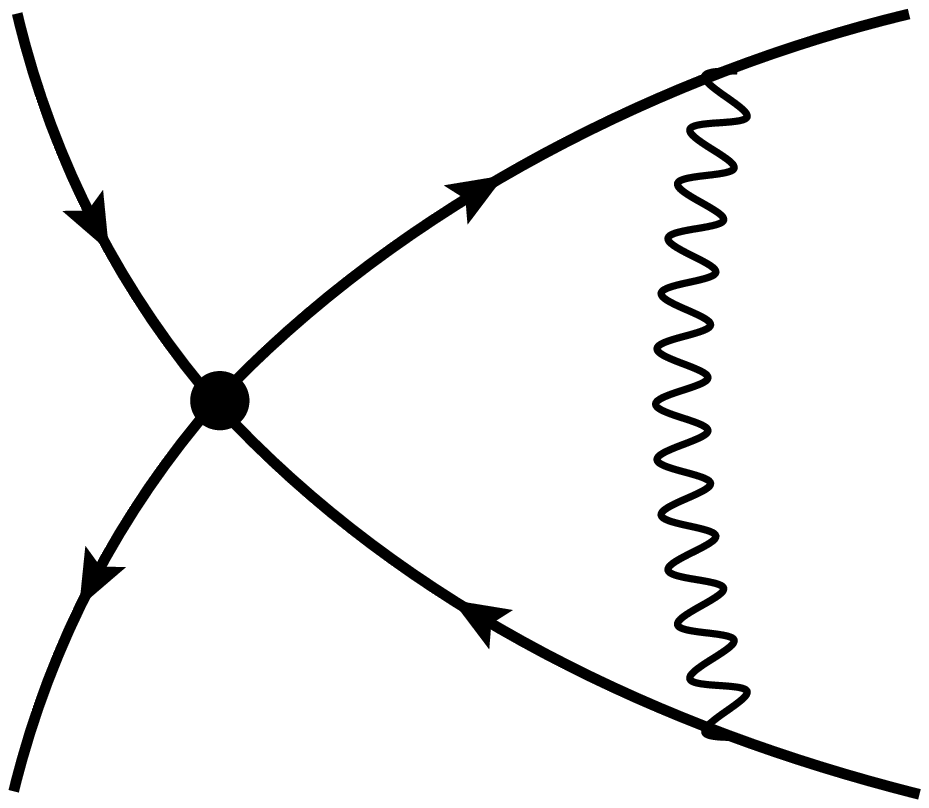}\\[1mm]
(a)&\hspace*{05mm}(b)&\hspace*{05mm}(c)\\
\end{tabular}
\end{center}
\caption{\label{fig::1}   (a)  planar and  (b) nonplanar one-loop box
diagrams giving rise to the leading-power Glauber phase;  (c) an effective
theory diagram giving rise to the power-suppressed double-logarithmic
contribution. The black dot corresponds to the effective contact
${\cal O}(\tau)$ interaction in Eq.~(\ref{eq::Born}) and the symmetric
diagram is  not shown.}
\end{figure}

Let us now discuss the one-loop amplitude. The corrections enhanced by a
power of $\ln\tau$ can only be produced by the one-particle irreducible box
diagrams in Fig.~\ref{fig::1}(a,b). Hence, we do not consider the
factorizable correction where a gauge boson is emitted and absorbed by the
same fermion line. The double-logarithmic contribution comes from the
virtual momentum region where $l\ll \sqrt{s}$ and the virtual fermion or
antifermion  is close to its mass-shell \cite{Gorshkov:1973if}. Thus we can
use the same scaling rules for the components of $l$ as  for the components
of the momentum transfer $q$. Then it is sufficient to consider the eikonal
approximation for the (anti)fermion propagator
\begin{eqnarray}
S(l-p_{1,2})&\to& {\gamma_\mp\over  2l_\mp\pm l_\perp^2/\sqrt {s/2}\pm i\epsilon}\,,
\label{eq::prop}
\end{eqnarray}
and the static approximation for the Glauber gauge boson  propagator
$D_{\mu\nu}(l)\to-g_{+-}/l^2_\perp$.  Let us first  briefly discus the
corresponding leading-power contribution. In this case the   ${\cal
O}(1/\sqrt{s})$ terms in the eikonal propagators can be neglected and  each
gauge boson exchange is described by the leading term of Eq.~(\ref{eq::Born})
with the corresponding momentum transfer.  Then the planar box diagram
Fig.~\ref{fig::1}(a) is given by the amplitude ${\cal M}_v$ with a factor
\begin{equation}
{\alpha\over \pi}{it\over (2\pi)^2}
\int{{{\rm d}^2{l}_\perp}{{\rm d}l_+}{{\rm d}l_-}\over l^2_\perp(l-q)_\perp^2
\left(l_-+i\epsilon\right)\left(l_+-i\epsilon\right)}\,,
\label{eq::lpint}
\end{equation}
where $\alpha=g^2/(4\pi)$. The expression for the nonplanar diagram
Fig.~\ref{fig::1}(b) differs only by the sign of $l_+$ and in the sum  the
antifermion propagators add up to $1/(2l_+-i\epsilon)-c.c.=i\pi\delta(l_+)$.
By symmetrization the fermion propagator   can be reduced to
$-i\pi\delta(l_-)$ in the same way. After trivial integration over the
light-cone components the total leading-power contribution is purely
imaginary and can be written as $i\phi{\cal M}_v$ with the Glauber phase
given by an infrared divergent transverse momentum integral
\begin{equation}
\phi=-{\alpha}\int{  {\rm d}^2l_\perp\over 2\pi}
{q_\perp^2\over l^2_\perp(l-q)^2_\perp}= -{\alpha}\ln\left({-t\over \mu^2}\right)\,,
\label{eq::phase}
\end{equation}
where $\mu$ is an infrared regulator. The Glauber  phase is known to
exponentiate so that the all-order  leading-power amplitude  is given by
$e^{i\phi}{\cal M}_v$  and has no $\ln\tau$ terms in any order of
perturbation theory \cite{Cheng:1970jk,Chang:1969by}.\footnote{This property
is violated by the light-by-light scattering contribution
\cite{Cheng:1970xm,Frolov:1970ij} discussed below.}

The calculation of the double-logarithmic one-loop ${\cal O}(\tau)$ amplitude
can be performed in a similar way. However, in this case the power suppressed
term in the eikonal propagators should be kept and for  one of the gauge boson
exchanges the contact ${\cal O}(\tau)$ part of the amplitude
Eq.~(\ref{eq::Born}) should be taken. The full effective theory one-loop
expression includes also the diagrams with a local ${\cal O}(g^2\tau)$ vertex
quadratic in the gauge field which accounts for the off-shell virtual
(anti)fermion contribution. Such diagrams however do not have the
double-logarithmic scaling  and are omitted. The  planar diagram is then
proportional to ${\cal M}_v+{\cal M}_a$ {\it i.e.} contributes only to the
same-helicity fermion scattering. The  corresponding coefficient is infrared
finite and with the double-logarithmic accuracy is given by the integral
\begin{equation}
-{\alpha\over \pi}{i\tau\over (2\pi)^2}
\int{{{\rm d}^2{l}_\perp}{{\rm d}l_+}{{\rm d}l_-}\over l_\perp^2
\left(l_-+l^2_\perp/\sqrt{2s}\right)\left(l_+-l^2_\perp/\sqrt{2s}\right)}
\,
\label{eq::spint}
\end{equation}
represented by the effective theory Feynman diagram in Fig.~\ref{fig::1}(c).
Note that in Eq.~(\ref{eq::spint}) we put $q=0$ and this expression is valid
only for $l\gg q$, which is sufficient for the calculation of the
double-logarithmic terms. Integration over $l^+$  in Eq.~(\ref{eq::spint})
can be performed by using Cauchy theorem and taking the residue of the
antifermion propagator pole, which gives $l_+={l}_\perp^2/ (\sqrt{2s})$ and
$l_->0$. Then the remaining integral has the double-logarithmic scaling in
the interval ${l}_\perp^2/ \sqrt{s}<l_-<\sqrt{s}$ and
$q_\perp^2<{l}_\perp^2<s$. Thus Eq.~(\ref{eq::spint}) with the
double-logarithmic accuracy can be rewritten as follows
\begin{equation}
{\tau\alpha\over 2\pi}\int^s_{q_\perp^2}{{{\rm d}{l}^2_\perp}
\over l_\perp^2}\int^{\sqrt{s}}_{l_\perp^2/\sqrt{s}}{{{\rm d}l_-} \over
l_-}=\tau z\int^1_0{\rm d}\xi\int^\xi_0{\rm d}\eta={\tau z\over 2}\,,
\label{eq::logint}
\end{equation}
where we introduce the  variables  $\xi=\ln(l_\perp^2/s)/\ln\tau$,
$\eta=\ln(l_-/\sqrt {s})/\ln\tau$, and $z={\alpha\over 2\pi} \ln^2\!\tau$. As
it was discussed before the nonplanar $u$-channel  diagram is proportional to
${\cal M}_v-{\cal M}_a$ and contributes  only to the opposite helicity
fermion scattering. In the sum of the planar and nonplanar contributions the
axial part cancel  and the total one-loop amplitude takes the form
\begin{equation}
{\cal M}^{(1)}=\left(i\phi+\tau z+{\cal O}(\tau^2)\right)
{\cal M}^{(0)}\,,
\label{eq::oneloopamp}
\end{equation}
in agreement with the expansion of the exact result for the chiral
amplitudes (see, {\it e.g.} \cite{Kuhn:2001hz}).

\begin{figure}[t]
\begin{center}
\begin{tabular}{ccc}
\includegraphics[width=1.8cm]{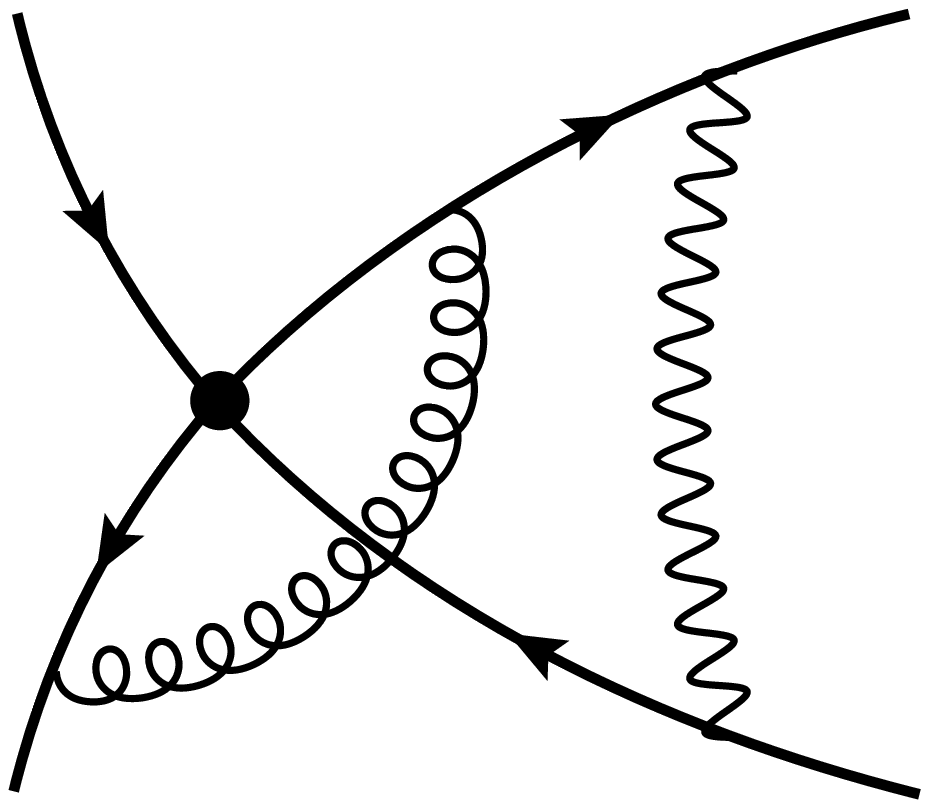}&
\hspace*{05mm}\includegraphics[width=1.8cm]{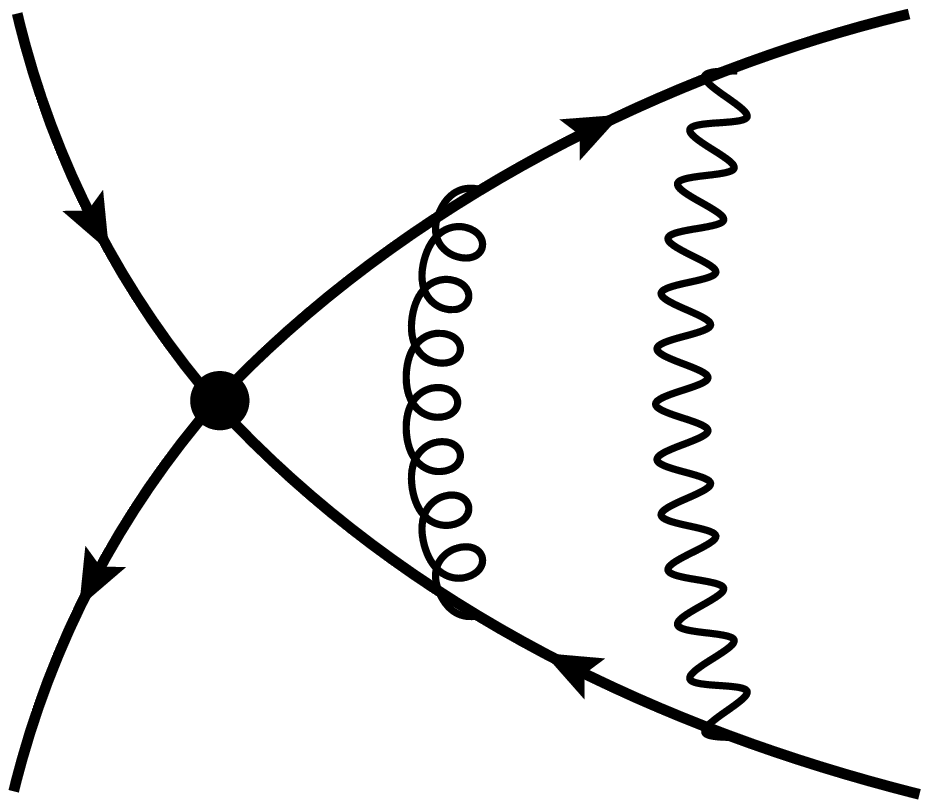}&
\hspace*{05mm}\includegraphics[width=1.8cm]{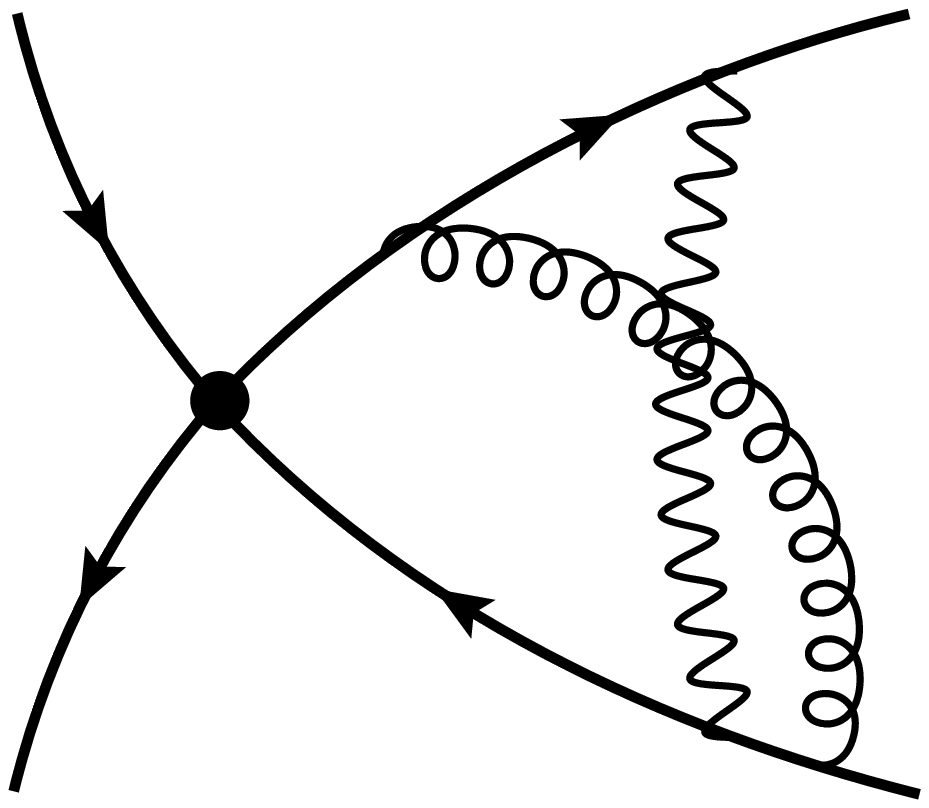}\\[1mm]
(a)&\hspace*{05mm}(b)&\hspace*{05mm}(c)\\
\end{tabular}
\end{center}
\caption{\label{fig::2}  Two-loop diagrams generating the double-logarithmic
power-suppressed corrections. The loopy line corresponds to the leading-power
Glauber gauge boson. The symmetric diagrams are not shown.}
\end{figure}

In the  near-forward scattering the charged particles almost do not accelerate
and the usual double-logarithmic corrections associated with the soft gauge
boson emission are suppressed  \cite{Gorshkov:1973if}.  Thus the higher-order
double-logarithmic corrections are due to multiple  leading-power Glauber gauge
boson exchanges which produce infrared-finite terms of the form
$(\alpha\ln^{2}\!\tau)^n$. To determine  the factorization structure of these
corrections let us consider the two-loop power-suppressed contribution obtained
by dressing the one-loop diagrams with a Glauber gauge boson of the momentum
$l'$. The corresponding effective theory expression can be directly obtained  by
expanding the  full theory diagrams in $l$ and $l'$ and keeping only the power
suppressed terms  linear in $l_\pm\sim \tau$ and quadratic in
$l_\perp\sim\tau^{1/2}$ which give rise to the effective contact interaction as
in Fig.~\ref{fig::1}c. Then in the sum of all the diagrams one can separate the
part proportional to  the product $\delta(l'_-)\delta(l'_+)$. As for  the
leading-power contribution it gives the infrared-divergent logarithmic Glauber
phase Eq.~(\ref{eq::phase}), which factors out with respect to the one-loop
${\cal O}(\tau)$ amplitude. The remaining contribution reduces to  the effective
theory diagrams with the characteristic structure shown in Fig.~\ref{fig::2}. In
the sum over all the permutations of the gauge boson vertex along the eikonal
antifermion line, Fig.~\ref{fig::2}(a-c), the dependence on the light-cone
component of its momentum $l'$ can be factored out into $\delta(l'_+)$. At
the same time the Clauber gauge boson is  attached only  to the virtual fermion
line carrying the loop momentum $l$ and  the integration over $l'_-$ remains
logarithmic. The logarithmic integration intervals can be easily identified as
$q_\perp^2<{l'}_\perp^2<l_\perp^2$,  ${l}_-<l'_-<\sqrt {s}$ and after
integration over $l'_+$ the second loop results in a factor
\begin{equation}
{2\alpha\over \pi} \int^{{l}^2_\perp}_{q_\perp^2}{{{\rm d}{l'}^2_\perp}
\over {l'}_\perp^2}\int^{\sqrt{s}}_{l_-}{{{\rm d}l'_-} \over
l'_-}=4z \int^1_\xi{\rm d}\xi'\int^\eta_0{\rm d}\eta'=4z\eta(1-\xi)\,.
\label{eq::logintp}
\end{equation}
It is straightforward to check that for an arbitrary number $n$ of the
leading-power Glauber gauge bosons with the momenta $l'_i$ the factorization
structure remains the same: the Glauber phase exponentiates  and the
double-logarithmic contribution is determined by the sum of the diagrams with
all the leading-power vertices attached to the virtual fermion line carrying the
loop momentum $l$, as in Fig.~\ref{fig::2}. Again the dependence on ${l'_i}_+$
factors out to $\delta({l'_i}_+)$ and the logarithmic integration intervals are
$q_\perp^2<{l_i'}_\perp^2<l_\perp^2$ and ${l}_-<{l'_1}_-<\ldots<{l'_n}_-<\sqrt
{s}$, {\it i.e.} the light-cone components are ordered along the eikonal fermion
line. The integration over $l'_i$ then gives ${\left(4z\eta(1-\xi)\right)^n/n!}$
and after summation over $n$ we get an exponential factor $e^{4z\eta(1-\xi)}$ to
be inserted  into the remaining loop integral Eq.~(\ref{eq::logint}). Thus in
the leading (double) logarithmic approximation the scattering amplitude reads
\begin{equation}
{\cal M}^{LL}=\tau z\,g(2z){\cal M}^{(0)}\,,
\label{eq::llamp}
\end{equation}
where
\begin{equation}
g(z)=2\int^1_0{\rm d}\xi\int^\xi_0{\rm d}\eta\,
e^{2z\eta(1-\xi)}
={}_2F_2\left(1,1;{3/2},2;{z/2}\right)\,
\label{eq::g}
\end{equation}
is the  generalized hypergeometric function with the  Taylor series
expansion $g(z)=1+z/6+z^2/45+\ldots$.  The ${\cal O}(z^2)$ contribution to
Eq.~(\ref{eq::llamp}) agrees with the small-angle expansion of the two-loop
result for  Bhabha scattering \cite{Bern:2000ie, Penin:2005kf,Penin:2005eh},
which is a nontrivial check of our calculation. It is interesting to note
that exactly the same function $g(z)$ appears in the analysis of the
double-logarithmic asymptotic behavior of the mass-suppressed QED and QCD
amplitudes in the Sudakov limit. Though it is fairly easy to notice a similar
factorization structure in both cases, such a universality is rather
surprising since in Sudakov limit the double-logarithmic corrections are
associated with the exchange of the on-shell  rather than Glauber gauge bosons.

\begin{table}[t]
  \begin{ruledtabular}
    \begin{tabular}{c|c|c|c|c}
      & i &  ii & iii & iv \\
     \hline
     $\gamma$  &
     ${\alpha\over 2\pi}\ln\left|{s/t}\right|$&
     $\left({2\alpha\over \pi}\right)^{1/2}$&
     $1+{11\pi\over 36}\alpha^2$&
     $1+{4\ln2\over \pi} N_c\alpha_s$
      \\
    \end{tabular}
    \end{ruledtabular}
    \caption{\label{tab::1}  The  exponent $\gamma$ defined in the text for
    (i) the subleading-power scattering, Eq.~(\ref{eq::llasy});  (ii)
    electron-to-muon pair forward annihilation  amplitude \cite{Gorshkov:1966ht};
    (iii) QED Regge cut  contribution \cite{Cheng:1970xm,Frolov:1970ij}; (iv)
    BFKL  pomeron contribution     \cite{Kuraev:1977fs,Balitsky:1978ic}.  }
\end{table}

The function $g(z)$ has the following asymptotic behavior at $z\to\infty$
\begin{equation}
\quad g(z)\sim \left({2\pi e^{z}\over z^{3}}\right)^{1/2}\!\!.
\label{eq::gasymp}
\end{equation}
Thus in  Regge limit we have
\begin{equation}
{\cal M}^{LL}\sim
{\pi/\sqrt{2 \alpha}\over \ln\left|s/t\right|}
\left|{s\over t}\right|^{-1+{\alpha\over 2\pi}\ln\left|{s/t}\right|}
{\cal M}^{(0)}\,,
\label{eq::llasy}
\end{equation}
which is the main result of this Letter.  To characterize the asymptotic
behavior of the  amplitudes at high energy  it is convenient to introduce an
exponent $\gamma$  so that ${\cal M}\sim s^\gamma$ for $s\to\infty$. Since
${\cal M}^{(0)}\sim s$, our result Eq.~(\ref{eq::llasy}) corresponds to
$\gamma={\alpha\over 2\pi}\ln\left|{s/t}\right|$. The values of  $\gamma$ for
a number of different amplitudes and gauge groups are collected  in
Table~\ref{tab::1}. In particular it includes the results for the
leading-power amplitudes corresponding to the contribution of the rightmost
singularities  in the  complex angular momentum plane: the light-by-light
scattering induced Regge cut in QED \cite{Cheng:1970xm,Frolov:1970ij} and the
BFKL pomeron Regge pole in QCD \cite{Kuraev:1977fs,Balitsky:1978ic}. The
common feature of all these examples is that $\gamma>0$ and the radiative
corrections result in the asymptotic enhancement of the amplitudes. At the
same time only for the subleading-power scattering the exponent depends on
$\tau$, which is characteristic to the double-logarithmic corrections. By
contrast the leading-power effects are single-logarithmic and give
energy-independent $\gamma$. On the other hand the QED forward annihilation
amplitude \cite{Gorshkov:1966ht} is power-suppressed and does get the
double-logarithmic corrections. However, at $z\to\infty$ this amplitude
becomes an exponential function of $z^{1/2}$ rather than $z$  so that  the
corresponding exponent $\gamma\sim \sqrt{\alpha}$ is nonanalytic in coupling
constant but does not depend on $\tau$ and is consistent with the square root
branch point in the complex angular momentum plane.

We can apply the above result to describe  the power-suppressed effect in the
color-singled channel of quark-aniquark scattering in QCD. In this case  in
the expression for $\gamma$ one should use $\alpha=\left({N_c^2-1\over
4N_c^2}\right)^{1/2}\!\!\!\alpha_s$,  where $N_c=3$ is the number of colors,
$\alpha_s$ is the strong coupling constant, and the prefactor arises from
projecting on the color-singlet configuration. In principle in the
color-singlet channel one should retain only the contribution of even number
of gluons {\it i.e.} only  even powers of $z$ in the Taylor series for
$g(z)$. This however does not change the value of  $\gamma$ in the asymptotic
formula. Though this result clearly does not give a complete account of
subleading-power effects in Regge limit of QCD, it can be considered as a
power correction to the BFKL pomeron contribution to the elastic parton
scattering \cite{Mueller:1992pe}.

In the high-energy limit for a given $\gamma$ the total cross section has the
following scaling $\sigma\sim s^{2(\gamma-1)}$. For  $\gamma>1$ such an
asymptotic behavior violates the Froissart  bound  \cite{Froissart:1961ux}
and hence the S-matrix  unitarity.  This constitutes the unitarity problem of
the perturbative Regge analysis, which gives $\gamma>1$  both in QED and QCD,
{\it cf.} Table~\ref{tab::1}. Though a precise mechanism of unitarity
restoration is not yet known,  our result sets a {\it perturbative} bound on
the energy scale where the Regge analysis of the leading-power amplitudes can
be trusted. Indeed, for $\left|\ln\tau\right|\sim 1/\alpha$ or more precisely for
\begin{equation}
\left|{s\over t}\right|\approx e^{2\pi/\alpha}\,
\label{eq::scale}
\end{equation}
the formally power-suppressed contribution to the cross section becomes
comparable to the leading-power one and the small-angle expansion breaks
down. Interestingly this is roughly the scale at which the resummation
of the single-logarithmic corrections responsible for the Regge
behavior of the leading-power amplitudes becomes mandatory.

\begin{acknowledgments}
I  would like to thank  Kirill Melnikov for discussions and collaboration and
Tao Liu for important cross-checks. This work  is supported in part by NSERC
and  Perimeter Institute for Theoretical Physics.
\end{acknowledgments}



\begin{thebibliography}{99}



\bibitem{Cheng:1970jk}
  H.~Cheng and T.~T.~Wu,
  Phys.\ Rev.\  {\bf 186}, 1611 (1969).


\bibitem{Chang:1969by}
  S.~J.~Chang and S.~K.~Ma,
  Phys.\ Rev.\  {\bf 188}, 2385 (1969).

\bibitem{Cheng:1970xm}
  H.~Cheng and T.~T.~Wu,
  Phys.\ Rev.\ D {\bf 1}, 2775 (1970).


\bibitem{Frolov:1970ij}
  G.~V.~Frolov, V.~N.~Gribov and L.~N.~Lipatov,
  Phys.\ Lett.\  {\bf 31B}, 34 (1970).


\bibitem{Lipatov:1976zz}
  L.~N.~Lipatov,
  Sov.\ J.\ Nucl.\ Phys.\  {\bf 23}, 338 (1976)
  [Yad.\ Fiz.\  {\bf 23}, 642 (1976)].


\bibitem{Kuraev:1977fs}
  E.~A.~Kuraev, L.~N.~Lipatov and V.~S.~Fadin,
  Sov.\ Phys.\ JETP {\bf 45}, 199 (1977)
  [Zh.\ Eksp.\ Teor.\ Fiz.\  {\bf 72}, 377 (1977)].

\bibitem{Balitsky:1978ic}
  I.~I.~Balitsky and L.~N.~Lipatov,
  Sov.\ J.\ Nucl.\ Phys.\  {\bf 28}, 822 (1978)
  [Yad.\ Fiz.\  {\bf 28}, 1597 (1978)].

\bibitem{Collins:2009} P.~D.~B. Collins, {\it An introduction to
   Regge theory and high-energy physics},
   Cambridge Monographs on Mathematical Physics,
   Cambridge University Press, Cambridge UK, 2009.

\bibitem{Fadin:1998py}
  V.~S.~Fadin and L.~N.~Lipatov,
  Phys.\ Lett.\ B {\bf 429}, 127 (1998).

\bibitem{Gorshkov:1966ht}
  V.~G.~Gorshkov, V.~N.~Gribov, L.~N.~Lipatov and G.~V.~Frolov,
  Sov.\ J.\ Nucl.\ Phys.\  {\bf 6}, 95 (1968)
  [Yad.\ Fiz.\  {\bf 6}, 129 (1967)].


\bibitem{Penin:2014msa}
  A.~A.~Penin,
  Phys.\ Lett.\ B {\bf 745}, 69 (2015), Erratum: [Phys.\ Lett.\ B {\bf 771}, 633 (2017)].


\bibitem{Melnikov:2016emg}
  K.~Melnikov and A.~Penin,
  JHEP {\bf 1605}, 172 (2016).


\bibitem{Penin:2016wiw}
  A.~A.~Penin and N.~Zerf,
  Phys.\ Lett.\ B {\bf 760}, 816 (2016),  Erratum: [Phys.\ Lett.\ B {\bf 771}, 637 (2017)].

\bibitem{Liu:2017vkm}
  T.~Liu and A.~A.~Penin,
  Phys.\ Rev.\ Lett.\  {\bf 119},  262001 (2017).


\bibitem{Liu:2018czl}
  T.~Liu and A.~Penin,
  JHEP {\bf 1811}, 158 (2018).

\bibitem{Glauber}
  R.~J.~Glauber, in {\em Lectures in Theoretical Physics}, edited by
  W. E. Brittin et al., Wiley-Interscience Inc., New York, 1959,  Vol. 1.


\bibitem{Gorshkov:1973if}
  V.~G.~Gorshkov,
  Usp.\ Fiz.\ Nauk {\bf 110} (1973) 45.

\bibitem{Kuhn:2001hz}
  J.~H.~Kuhn, S.~Moch, A.~A.~Penin and V.~A.~Smirnov,
  Nucl.\ Phys.\ B {\bf 616}, 286 (2001)
  Erratum: [Nucl.\ Phys.\ B {\bf 648}, 455 (2003)].


\bibitem{Sudakov:1954sw}
  V.~V.~Sudakov,
  Sov.\ Phys.\ JETP {\bf 3}, 65 (1956)
  [Zh.\ Eksp.\ Teor.\ Fiz.\  {\bf 30}, 87 (1956)].

\bibitem{Bern:2000ie}
  Z.~Bern, L.~J.~Dixon and A.~Ghinculov,
  Phys.\ Rev.\ D {\bf 63}, 053007 (2001).

\bibitem{Penin:2005kf}
  A.~A.~Penin,
  Phys.\ Rev.\ Lett.\  {\bf 95}, 010408 (2005).

\bibitem{Penin:2005eh}
  A.~A.~Penin,
  Nucl.\ Phys.\ B {\bf 734}, 185 (2006).

\bibitem{Mueller:1992pe}
  A.~H.~Mueller and W.~K.~Tang,
  Phys.\ Lett.\ B {\bf 284}, 123 (1992).

\bibitem{Froissart:1961ux}
  M.~Froissart,
  Phys.\ Rev.\  {\bf 123}, 1053 (1961).

\end{thebibliography}
\end{document}